# Mapping the aberrations of a wide-field spectrograph using a photonic comb

**JOSS BLAND-HAWTHORN,**[1,2,3] **JANEZ KOS,**[3] **CHRISTOPHER H. BETTERS,**[1,2,3] **GAYANDHI DE SILVA,**[3,4] **JOHN O'BYRNE,**[1,3] **ROB PATTERSON,**[5] **SERGIO G. LEON-SAVAL**[1,2,3]

[1] *Sydney Astrophotonic Instrumentation Labs, School of Physics, University of Sydney, NSW 2006, Australia*
[2] *Institute of Photonics and Optical Science, School of Physics, University of Sydney, NSW 2006, Australia*
[3] *Sydney Institute for Astronomy, School of Physics, University of Sydney, NSW 2006, Australia*
[4] *Australian Astronomical Observatory, 105 Delhi Rd, North Ryde, NSW 2113, Australia*
[5] *Anglo-Australian Telescope, Siding Spring Observatory, NSW 2357, Australia*

**Abstract:** We demonstrate a new approach to calibrating the spectral-spatial response of a wide-field spectrograph using a fibre etalon comb. Conventional wide-field instruments employed on front-line telescopes are mapped with a grid of diffraction-limited holes cut into a focal plane mask. The aberrated grid pattern in the image plane typically reveals n-symmetric (e.g. pincushion) distortion patterns over the field arising from the optical train. This approach is impractical in the presence of a dispersing element because the diffraction-limited spots in the focal plane are imaged as an array of overlapping spectra. Instead we propose a compact solution that builds on recent developments in fibre-based Fabry-Perot etalons. We introduce a novel approach to near-field illumination that exploits a 25cm commercial telescope and the propagation of skew rays in a multimode fibre. The mapping of the optical transfer function across the full field is represented accurately (<0.5% rms residual) by an orthonormal set of Chebyshev moments. Thus we are able to reconstruct the full 4K×4K CCD image of the dispersed output from the optical fibres using this mapping, as we demonstrate. Our method removes one of the largest sources of systematic error in multi-object spectroscopy.

## 1. Introduction

A conventional approach to mapping the aberrations of a wide-field imaging or focal reducer optics is to illuminate a focal plane mask at the input with a lambertian continuum source. The mask is usually a 2D grid of diffraction limited circular holes cut to high precision[1]. The aberrated grid pattern in the image plane typically reveals centrosymmetric or n-fold symmetric (e.g. pincushion) distortion patterns over the field arising from a convolution of the optical transfer function of the entrance aperture and each optical element. The aberrations are revealed through the spatial shifts, elliptic distortions and rotations of the imaged spots of light. The 2D array of "point spread functions" (PSF) can then be analysed by fitting image moments to isolate the different families of aberrations.

This approach is impractical in wide-field spectroscopy because the continuum point sources in the focal plane are dispersed in the image plane. We propose a novel solution that builds on recent developments in fibre-based etalon cavities. Moreover, we demonstrate an ingenious approach to near-field illumination by using a commercially available OD 25cm Celestron and propagating skew rays in a multimode fibre.

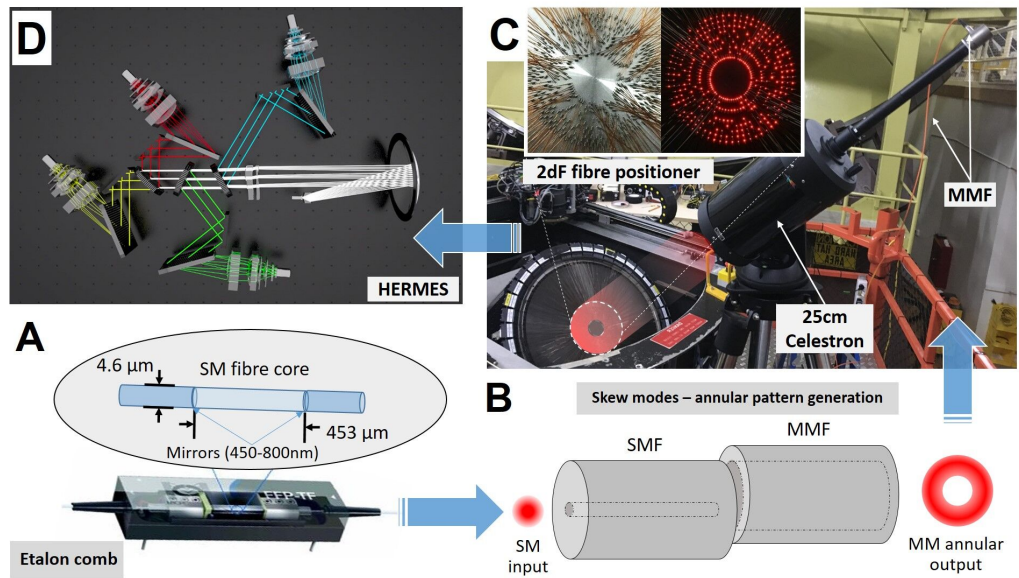

Fig. 1. The output of the fibre etalon (A) is fed into a large aperture multimode fibre (B) to excite skew rays. The annular signal is fed to a OD 25cm Celestron (C); the Cassegrain optics reflects a collimated annular beam with diameter 25cm. The central hole is to avoid the secondary mirror of the Celestron. The left inset shows the 400 fibres of the 2dF positioner configured into an annulus to receive the light. The right inset shows the fibres back-illuminated to reveal the annular pattern. The fibres feed the $15M HERMES multi-arm high resolution spectrograph (D); the linear dimension is 4 meters.

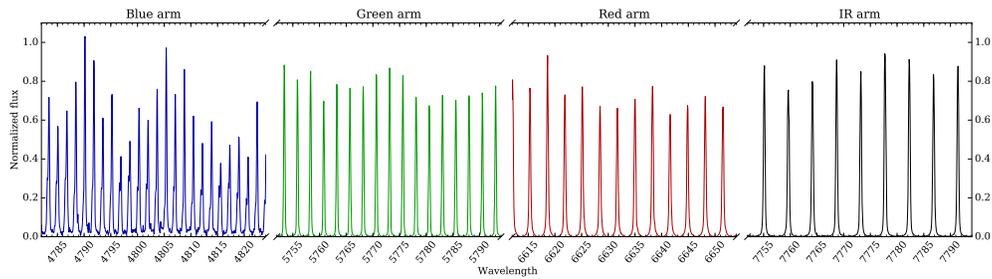

Fig. 2. Photonic comb spectra (subset) from a single fibre for all four arms (blue, green, red, IR) of the HERMES spectrograph. The blue arm shows double peaks because the fibre etalon becomes multimoded below 500nm. The tines are periodic in frequency.

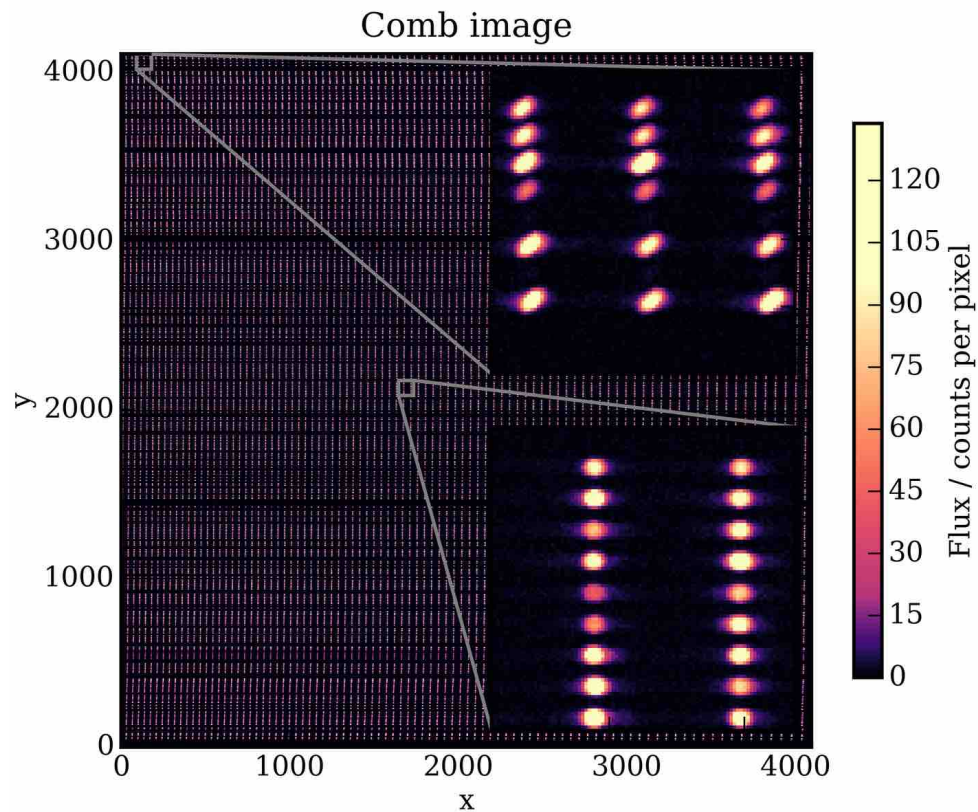

Fig. 3. Photonic comb spectra for the green arm imaged onto a 4K×4K CCD through 400 fibres placed along the lefthand vertical axis; each spot shows the PSF response for a given wavelength. The PSFs at large off-axis angles are distorted compared to the on-axis PSFs, mostly due to the wide-angle camera optics of the spectrograph.

## 2. Photonic comb and observational setup

We refer to photonic combs collectively as all devices that seek to produce a periodic output in frequency for the purposes of instrument calibration, both in terms of wavelength calibration and PSF mapping. This includes laser combs[2], fibre etalon combs[3], ring resonator combs[4] and so on. There are many ways to achieve this sort of output as described in recent reviews[5]. We reserve the term "frequency comb" for locking and stability, e.g. radio/optical frequency mixing, which is a key aspect of photonic combs being locked to a local oscillator or standard. This has been brilliantly exploited by atomic physicists and led to the 2005 Nobel Prize being awarded to Hänsch and Hall. We discuss the precision locking of fibre etalons in a companion paper.

Photonic combs came to attention in astronomy a decade ago in the context of improving on the iodine cell method for precision radial velocities to assist with the search for exoplanets. The first realizations of laser combs were unwieldy, unstable and expensive - big powerful lasers were fed into large etalon cavities in order to remove tines from the densely bunched optical frequency comb[6,7]. Compact ring resonators[4] and fibre etalons[8,9] have also been discussed as possible wavelength calibrators for astronomy although scientific results have not been presented to date. Our approach here is different in that we use a photonic comb to map the instrumental aberrations with the goal to understand them intuitively. Moreover, from prior knowledge of the spectra under study, we seek to demonstrate that the comb output can be used to construct an accurate model of the original image (reconstruction), thus demonstrating a complete understanding of the instrument's performance.

A spectrograph collects light from the telescope focal plane, disperses it and re-images the light onto a detector. The most challenging spectrographs are those that (a) operate at high spectral resolving power ($R = \lambda/\delta\lambda > 20{,}000$), (b) over a broad spectral band ($\Delta\lambda \sim \lambda$), and (c) over large solid angles. The push to pack more information onto ever-increasing detector formats has led to optical designs with light rays that subtend large physical angles (>10°) with respect to the optical axis. We consider all of these extremes in the context of the $15M HERMES multi-object spectrograph at the Anglo-Australian Telescope (AAT 3.9m), Siding Spring Observatory[10]. This multimetre-scale, floor-mounted workhorse spectrograph uses beam splitters to divide up the incoming light into four spectral bands (blue, green, red, IR). The instrument has a nominal spectral resolving power of $R \approx 28{,}000$ across the full range. The fibre etalon was made to order with a finesse $N_R \approx 40$ (see below) coated over the range 450-800 nm, a physical gap with spacing $L \approx 453$ microns and wavelength-dependent refractive index $\mu \approx 1.46$. We illuminate the fibre etalon with light from a Koheras "superK" supercontinuum source.

Table 1: Basic differences between commonly used moments in image analysis.

|  | Image moments | Zernike moments | Chebyshev moments | Legendre moments |
|---|---|---|---|---|
| Orthogonal |  | ✓ | ✓ | ✓ |
| Exact | ✓ |  | ✓ |  |
| No coordinate transformation | ✓ |  | ✓ |  |
| Easy reconstruction |  | ✓ | ✓ | ✓ |
| Intuitive low-order moments | ✓ | ✓ |  |  |

We allow the output light of our fibre etalon photonic comb (Fig. 1A) to illuminate the secondary mirror of a commercially available Celestron (OD 25cm). The reflected light on output from the primary mirror is now collimated to mimic the signal from space (Fig. 1C). But most of the Gaussian beam, however, is blocked by the central obstruction within the Celestron. To sidestep this problem, we feed the fibre etalon light at an offset position (Fig. 1B) to the front face of a multimode fibre (MMF). This excites skew rays in the MMF such that most of the output light is now in a fat annulus which bypasses the central obstruction after reflection. The MMF preserves the input angle but azimuthally scrambles the signal. We now arrange for the 400 fibres to be positioned by the robot into a broad annulus to match the MMF output (Fig. 1C). Illuminating the optics with our comb produces the four spectra across four arms in Fig. 2; the output of a single fibre is shown. The spacing of the comb peaks (free spectral range, FSR) gets longer with increasing wavelength, specifically 0.25nm at 570 nm and 0.456nm at 780 nm. For a fixed etalon spacing, the FSR in wavelength scales as $\lambda^2/2\mu L$ where $\mu$ is the refractive index of the glass spacer gap. Our design finesse[11] is the minimum to ensure the profiles are in the Lorentzian rather than the Airy limit to ensure a clean separation of the peaks and appropriate for minimizing defect finesse.

## 3. Observations and analysis

For brevity, we concentrate on one spectrograph arm (green) but the results for all arms are similar. Each of the four CCD images from HERMES (Fig. 1D) contains tens of thousands of PSF sub-images. There is a slight overlap between the fibres (cross-talk) which gets worse towards the edges. A traditional arc lamp can produce many peaks but they vary significantly in power, can lead to gaps in the wavelength dimension and produce blended peaks which are difficult to disentangle. The photonic comb peaks have an intrinsic, resolved Lorentzian shape which must be deconvolved before the peaks are a fair representation of the PSF. We use the Richardson-Lucy algorithm to remove the Lorentzian profile. Because the Lorentzian profile is well sampled, minimal noise and PSF distortion is introduced at this step.

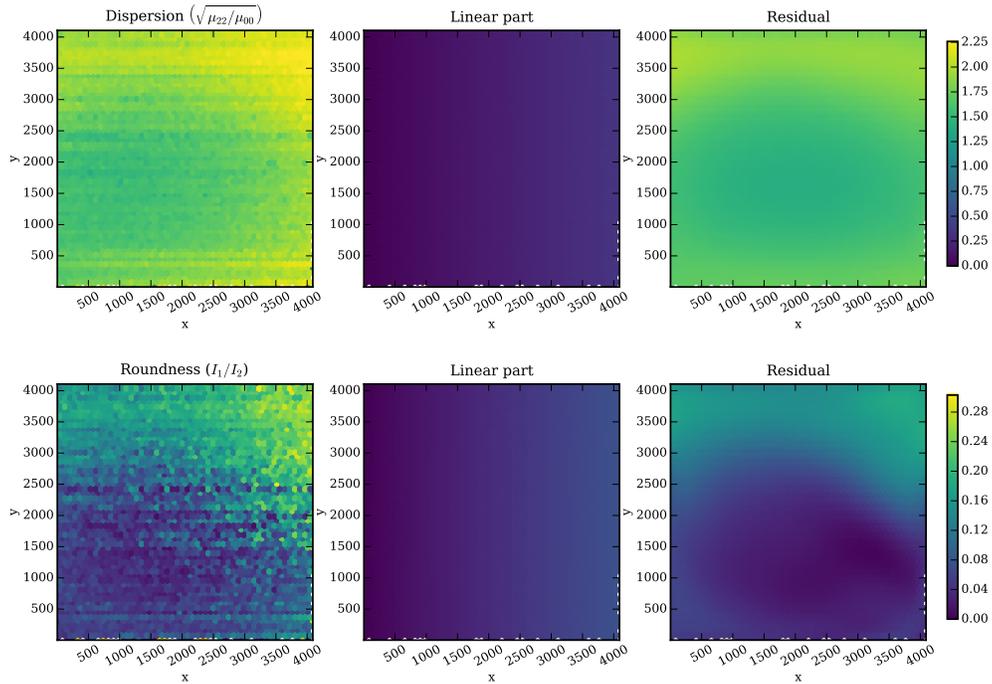

Fig. 4. Image moment analysis of the photonic comb spots across the full CCD for the green arm. (Top) Dispersion of the line profile in pixels showing that the PSF is well sampled everywhere. The PSF spots broaden at large off-axis angles from the field centre (left) with an additional weak gradient from left to right (middle); once the gradient is removed, the centro-symmetric broadening of the spots is more apparent (right). (Bottom) Roundness image defined as ratio of major to minor axis minus 1. The spots become elliptical at large off-axis angle; the decomposition is as above. The rightmost images are smoothed to remove the fibre-to-fibre variations.

In Fig. 3, the off-axis spectra are strongly aberrated compared to the on-axis spectra. This is the aberration which we seek to map over all four arms of the spectrograph with a view to removing the fibre-to-fibre dependence. This is an unsolved problem in applied science. The PSF peaks are separated in the spatial direction by 8 pixels or more. In order to analyze each peak individually and prevent the neighboring peaks from interfering, we cut out an area of

7×7 pixels around each peak and analyze these 49 pixels without any knowledge of the rest of the image. Orthogonal moments are best suited for this job as they have minimal redundancy and are ideal for rapid data reconstruction. They are, however, not the best choice when properties of the PSF are to be understood intuitively. We therefore analyze the PSF variation in two ways: image moments (see below) for the purpose of the visualizing optical aberrations; and discrete Chebyshev moments[12] for the purpose of image reconstruction. Table 1 shows the difference between image moments, discrete Chebyshev moments, and other commonly used moments.

## 4. Image moments

For a 2-dimensional image $f(x,y)$ of size $N \cdot N$ the image moments of order $p,q$ are defined as:

$$T_{pq} = \sum_{x=0}^{N-1}\sum_{y=0}^{N-1} x^p y^q f(x,y). \tag{1}$$

We use central image moments, which are invariant to translation, defined as

$$\mu_{pq} = \sum_{x=0}^{N-1}\sum_{y=0}^{N-1} (x-\bar{x})^p (y-\bar{y})^q f(x,y). \tag{2}$$

Here we do not need to know the position of the PSF peak with subpixel accuracy; the positional distortions are not needed in our current analysis. The image moments have a simple physical interpretation:

- $\mu_{00}$ = integrated power of the PSF peak,
- $\mu_{22}/\mu_{00}$ = second moment about the peak (i.e. related to variance),
- $\mu_{33}/\mu_{00}$ = skewness of the peak,
- $\mu_{44}/\mu_{00}$ = kurtosis of the peak,
- $\frac{1}{2}\arctan\left(\frac{2\mu_{11}/\mu_{00}}{\mu_{20}/\mu_{00} - \mu_{02}/\mu_{00}}\right)$ = rotation of the dominant axis of the peak.

We also make use of two Hu moments[13]:

$$\begin{aligned}I_1 &= \eta_{20} + \eta_{02} \\ I_2 &= (\eta_{20} - \eta_{02})^2 + 4\eta_{11}^2\end{aligned} \tag{3}$$

where

$$\eta_{pq} = \frac{\mu_{pq}}{\mu_{00}^{[1+(p+q)/2]}} \tag{4}$$

and $I_1/I_2$ is associated with the "roundness" of the image. Unfortunately, there is no easy way to reconstruct the original image from the image moments[14] as they are not orthogonal, although this is feasible from a smaller number of moments. Ultimately, reconstruction is desirable for the most exacting work when the goal is to match the many spectra with a spectral database, say, in order to make detailed comparisons. The ability to match a given spectrum to a library of spectra should not depend on the degree of instrumental aberration within the determined errors.

The image moment analysis is presented in Fig. 4. The variation in the width of the comb peaks (dispersion) shows a slow degradation in the spectral resolving power as a function of off-axis angle. This was expected from the Zemax analysis of the HERMES instrument for all four bands.[10] In concert with the changing resolution, the PSFs become progressively more elliptical at larger off-axis angle. In the interests of brevity, we refrain from showing the higher order moments which also show slow trends across the imaged field.

## 5. Chebyshev moments

The photonic comb data allow us to reconstruct the observed data to high accuracy assuming the source spectrum is well known. In order to demonstrate this, we observed the twilight sky on an earlier telescope run in order to detect the fading sunlight through each of the fibres. Ideally, the twilight data and the comb data would have been taken on the same night; in fact, they were taken 6 months apart. By definition, the twilight spectra are intrinsically identical except for the deleterious effects of instrumental aberration. Here we use a discrete classical Chebyshev polynomial[12] of order $n$ expressed in terms of binomial coefficients[14] and defined as:

$$t_n(x) = n! \sum_{k=0}^{n} (-1)^{n-k} \binom{N-1-k}{n-k} \binom{n+k}{n} \binom{x}{k} \quad (5)$$

where $N$ (=7 in this work) is the size of one dimension of a PSF image in pixels. This constrains the number of possible discrete polynomials to $n = 0, 1, 2, …, N-1$. The following condition must also be satisfied:

$$\sum_{x=0}^{N-1} t_p(x) t_q(x) = \rho(p,N) \delta_{pq}, \quad 0 \leq p, \ q \leq N-1 \quad (6)$$

where

$$\rho(p,N) = \sum_{x=0}^{N-1} [t_p(x)]^2 = \frac{N(N^2-1)(N^2-2^2)\cdots(N^2-p^2)}{2p+1} \quad (7)$$

The Chebyshev moment of order $p,q$ of an image with size $N \times N$ and intensity $f(x,y)$ is defined as:

$$T_{pq} = [\rho(p,N), \rho(q,N)]^{-1} \sum_{x=0}^{N-1} \sum_{y=0}^{N-1} t_p(x) t_q(y) f(x,y). \quad (8)$$

The inverse transformation is given by

$$f(x,y) = \sum_{m=0}^{N-1} \sum_{n=0}^{N-1} T_{mn} t_m(x) t_n(y). \quad (9)$$

In practice, we sum up to $N_{max}$ (≈3-5) where $N_{max} < N-1$. The discrete classical Chebyshev polynomial does not use numerical approximations which means we can afford to calculate as many moments (for arbitrarily large $p$ and $q$ up to $N-1$) as desired. This is not true for Zernike or Legendre moments where high order moments suffer from numerical errors, depending on the numerical precision used in the calculations. Chebyshev moments also require no coordinate transformation and work in the original space of the image. Most other base functions require a normalization of the image into a uniform range, usually [-1:1]. Since we are dealing with image sizes of only 7×7 pixels, the base functions can be tabulated for all possible combinations of $n$, $x$, and $N$ and queried rapidly which significantly reduces the computational time. It has been shown[15] that Chebyshev moments are less sensitive to noise, an important feature when dealing with astronomical data.

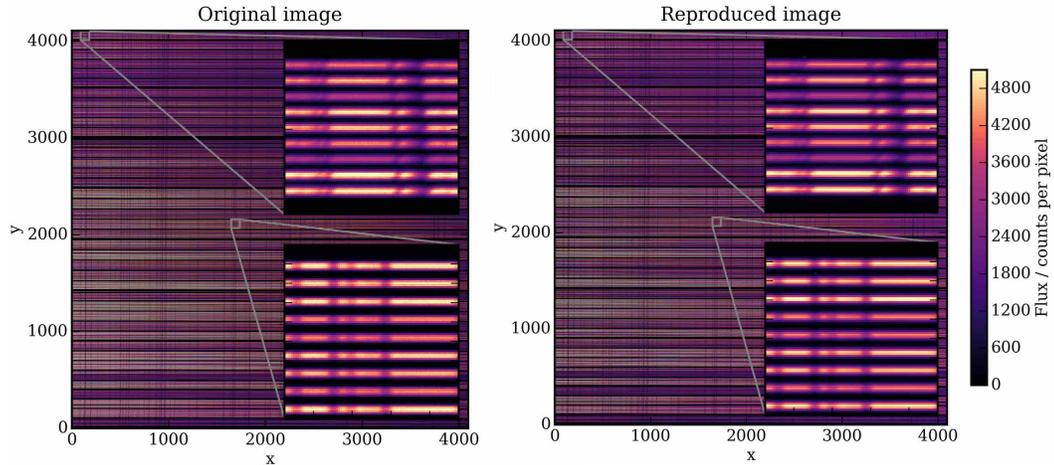

Fig. 5. (Left) Original spectral data of the twilight sky for the green arm. Note the on-axis spectra have well defined deep absorption lines, and the off-axis spectra are more aberrated. (Right) Reconstruction of the spectral data after convolving the solar spectral template with the photonic comb mapping function. The residual is shown in Fig. 6 below.

The reconstruction analysis is shown in Figs. 5 and 6. A small section of the solar (twilight) spectrum is shown in Fig. 5 (left) and its reconstruction in Fig. 5 (right). The results are excellent for on-axis spectra with <0.2% residuals within the spatial FWHM, <0.5% residuals averaged over the full fibre. Because of the uncalibrated cross-talk, the residuals climb to ≈1% for the most extreme off-axis spectra (Fig. 6). Our analysis reveals that the HERMES spectrograph has very low internal light scatter and ghost reflections (<0.5%) although measurable cross-talk (i.e. few percent) between fibres. This is the cause of the increased error at the edge of the field, combined with the strong aberrations here. In future work, we will illuminate interleaved sets of fibres separately to fully calibrate individual fibres. This will allow for a rigorous treatment of cross-talk between fibres leading to smaller residuals (Fig. 6) for the most extreme off-axis spectral aberrations.

## 6. The future

We have presented a completely general approach to calibrating the inherent distortions of a wide-field spectrograph. Another powerful application of photonic combs not explored here is their ability to map high order internal reflections, including both diametric and exponential ghosts. A demonstration of how such ghosts are revealed is given elsewhere[1].

With regard to the comb technology, the present limitation is the multimoded blue response of the fibre etalon and its relatively low power output. Bringing a small telescope into the field of view to illuminate many fibres at once is not a practical in the operation of a front-line (e.g. AAT 3.9m) telescope. The optimal approach to calibration is to illuminate a distant screen visible to the telescope. The fibre etalon incorporates a single-mode fibre with a 4-micron modal diameter. In order to greatly increase its étendue, we must consider other fibre technologies, specifically a photonic crystal fibre in order to generate "endlessly single mode" performance over a large aperture. Such a device is now under consideration for future use. The design will necessarily incorporate good single-mode performance down to the atmospheric cut-off at 330nm.

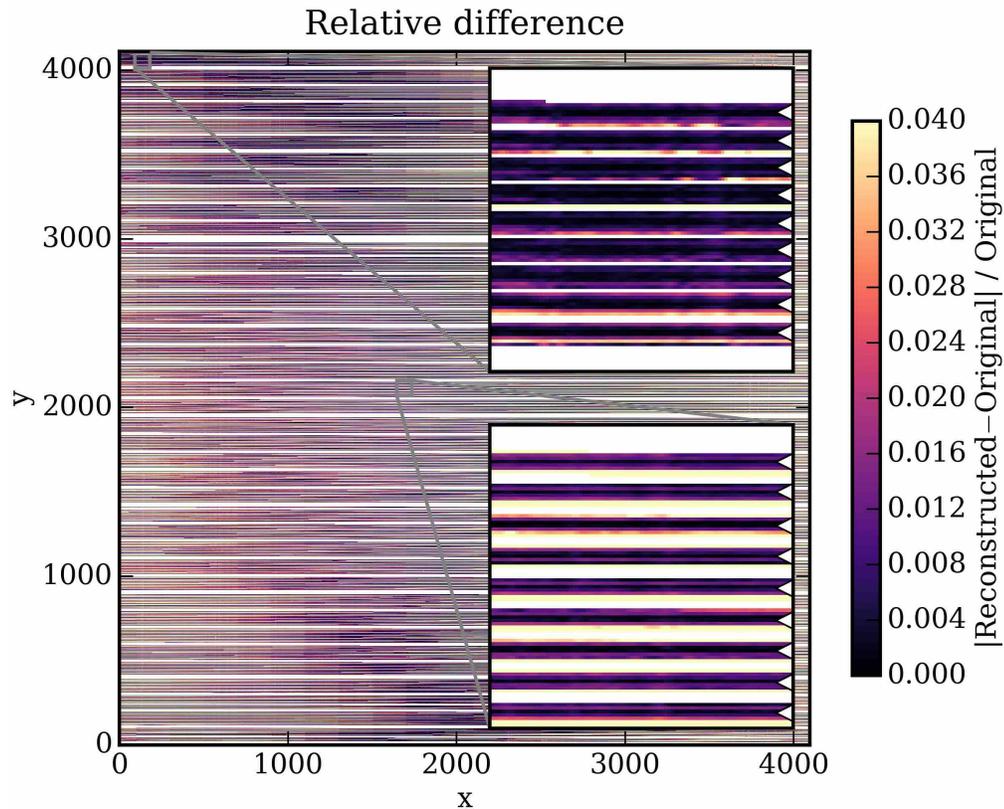

Fig. 6. The residual image when the reconstructed image is subtracted from the data in Fig. 5 divided by the data. The percentage errors are smoothed by the original PSF. The white pointers indicate the peak of the fibre signal. With our first attempt at reconstruction, we achieve better than 0.5% error on-axis with 1% errors at worst at the extreme off-axis position. The twilight and comb data were taken 6 months apart. The larger off-axis error is due to the combined effects of cross-talk between fibres and strong aberration; this will be corrected once the cross-talk is mapped.

The forward modeling analysis presented here is our first attempt at demonstrating the power of image reconstruction from using photonic combs. The method is so powerful that we believe future spectral analyses will treat multi-object spectra as a 2D image rather than extracting a set of 1D spectra treated individually, as we do today. In principle, this bypasses many of the existing problems of resampling, binning, tramline tracking and extraction. In general, *unlike* twilight observations, the objects observed in each fibre are not of the same spectral type. In other words, the many fibre outputs need to be reconstructed simultaneously and iteratively through forward modeling given the known optical distortion across the field. Ideally, we would treat all fibre outputs as a single entity because of their overlapping signal in the vertical direction. The cross-talk is of order a few percent of the total power; this is sufficiently high to compromise the reconstruction, particularly at the edges (Fig. 6). Thus *in toto* reconstruction will also resolve the long-standing and unresolved problem of cross-talk between fibres in wide-field spectrographs. These issues will be investigated in a future paper.

**Acknowledgments**

This research was carried out by members of the SAIL labs and supported through JBH's Laureate Fellowship from the Australian Research Council. We are indebted to J.G. Robertson and S.M. Croom for comments on the manuscript and acknowledge technical assistance at the telescope by the Australian Astronomical Observatory.